\documentclass{moriond}

\def\be{\begin{equation}}
\def\ee{\end{equation}}
\def\bea{\begin{eqnarray}}
\def\eea{\end{eqnarray}}

\newcommand{\Photo}{\vspace*{1cm}}

\begin{document}
\vspace*{3cm}
\title{SST-1M : Commissioning and Preliminary Observation Results}

\author{ Thomas Tavernier, for the SST-1M Collaboration}

\address{FZU - Institute of Physics of the Czech Academy of Sciences\\
Na Slovance 1999/2 182 00 Prague 8,
Czechia }

\maketitle\abstracts{
SST-1M is a single-mirror small size Cherenkov telescope prototype developed by a consortium among institutes in Switzerland, Poland, and the Czech Republic. With a 9.42 m² multi-segment mirror and a 5.6 m focal length, SST-1Ms have a broad 9 degree field of view and aim to detect gamma-rays spanning the energy range of 1 to 300 TeV. The DigiCam camera incorporates a compact Photo-Detector Plane comprising 1296 hexagonal silicon photomultiplier (SiPM) pixels and a fully digital readout and trigger system using a 250 MHz ADC.\\
Currently undergoing commissioning at the Ondrejov Observatory in the Czech Republic, two SST-1M telescopes are actively collecting data through observations of astrophysical gamma-ray sources. This presentation provides an overview of the telescope, camera design, and analysis pipeline, including evaluations of the instrument's responses. Preliminary results derived from ongoing observations are presented. A specific focus on data analysis of stereoscopic observations of the Crab Nebula provide insights on the telescope's sensitivity and performances.}

\section{Introduction}\label{sec:intro}

The SST-1M telescope was originally conceived to provide a cost-effective and 
performing solution for the implementation of  the small-sized telescope sub-array of the Cherenkov Telescope Array Observatory (CTAO) dominating the CTAO sensitivity above 1 TeV. Its mirror follows the Davies-Cotton concept and consists of 18 hexagonal facets, each 78 cm across, forming a spherical shape with a curvature radius of 11.2 m. This design maximizes the mirror's area while ensuring a precise optical point spread function (PSF) of less than 0.25°: 0.09° on-axis and 0.21° at a 4° off-axis angle. The telescope has a focal length of 5.6 m. The mirror area is 9.42 m without considering shadowing effects. Additionally, the mirror structure maintains an optical time spread of less than 0.84 nanoseconds at the focal plane.

The telescope's focal plane boats the DigiCam camera \cite{camH}. Its Photo-Detector Plane (PDP) is made of 1296 pixels, each utilizing a hexagonal light guide coupled with SiPM sensors. Equipped with a 250 MHz ADC, the camera ensures rapid signal processing and a  fully digital readout and trigger system. SiPM technology allow the camera to operate effectively in high Night Sky Background (NSB) conditions, thereby significantly enhancing the telescope's operational duty cycle without adopting protecting strategies for photosensors but by increasing the trigger threshold. The camera's fully digital trigger system utilizing FPGAs enables a versatile and adaptive approach, capable of implementing various triggering schemes simultaneously and adjusting per-pixel thresholds dynamically.

The stereo trigger of the SST-1M telescope relies on the CTA software array trigger (SWAT), ensuring coordinated detection of gamma-ray events across multiple telescopes. This system operates on trigger timestamps with nanosecond precision provided by the White Rabbit server that synchronizes internal clocks of both cameras.

In 2022, a pair of SST-1M telescopes were installed at the Ondřejov Observatory (500 m alt.) in the Czech Republic, positioned 152.5 m apart. These prototypes are currently undergoing commissioning and are actively collecting data through observations of both galactic and extra-galactic sources. 

%\begin{figure}
%\begin{minipage}{0.35\linewidth}
%\centerline{\includegraphics[width=0.9\linewidth]{fig/2tels.png}}
%\end{minipage}
%\hfill
%\begin{minipage}{0.65\linewidth}
%\centerline{\includegraphics[width=\linewidth]{fig/tel_drone2.png}}
%\end{minipage}
%\hfill
%\caption[]{Left panel: Photograph showcasing SST-1M telescopes. Right panel: Aerial view capturing the layout and surroundings of the telescopes}
%\label{fig:radish}
%\end{figure}

\begin{figure}
\center{
\includegraphics[width=.28\linewidth]{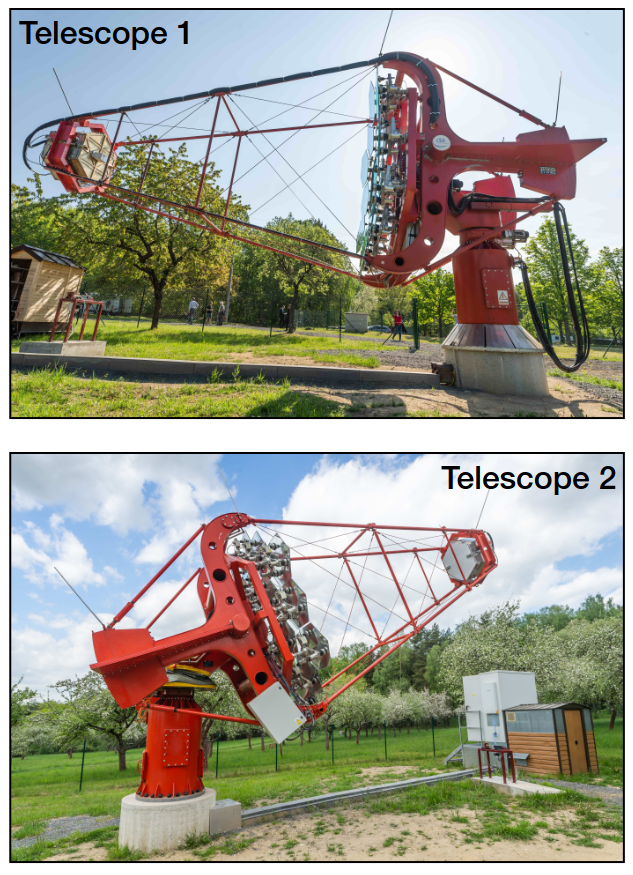}
\includegraphics[width=.59\linewidth]{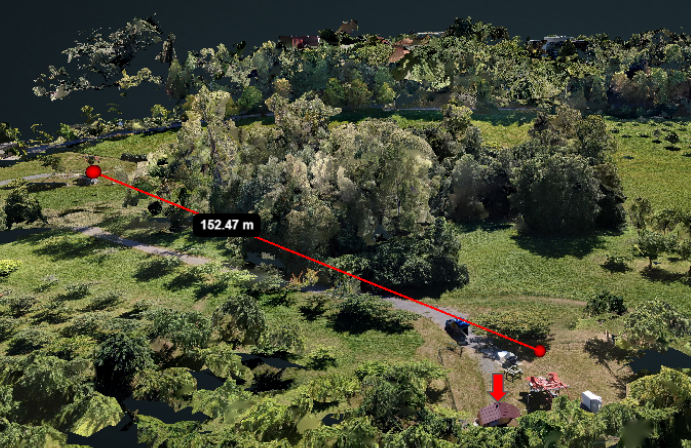}
}
\caption[]{Left panel: Photograph showcasing SST-1M telescopes. Right panel: Aerial view capturing the layout and surroundings of the telescopes}
\label{fig:radish}
\end{figure}

\section{Analysis pipeline}\label{sec:pipeline}
%is still currently in intensive development.
%inspired by lstchain and magic-cta-pipe

The \texttt{sst1mpipe} \cite{zen}$^,$ \cite{jj} analysis pipeline is designed to produce high level (DL3) data output following to the Gamma-ray Astronomy Data Format (GADF) standards, starting from the raw data produced by the instrument.

The pipeline is based on the ctapipe \cite{ctapipe} framework and encompasses all analysis  steps, including SiPM's response calibration, event image cleaning, gamma/hadron separation and reconstruction of the energy and direction of the primary particle utilizing random forests trained on Monte Carlo (MC) simulations. 
The pipeline also produces Instrument Response Functions (IRFs) in the GADF format based on MC simulations. This enables the production of higher-level data products, such as sky maps and spectra, trough the gammapy framework \cite{gammapy}.

The \texttt{sst1mpipe} pipeline is still currently in intensive development and the analysis of the commissioning datasets is crucial for fine tuning the MC simulations. It's essential to note that the results presented here are preliminary and subject to further refinement as the development of the pipeline progresses.

\section{Observation}\label{sec:obs}
\subsection{Crab Nebula}
The Crab Nebula is known to serve as a test beam in gamma-ray astronomy. Its exceptional luminosity and almost point like morphology makes it an ideal standard candle for calibration and comparative studies between different experiments.
We present here the results from the winter 2023/2024 Crab observation campaign,  started in November 2023 and concluded in March 2024. 
The data acquisition was ran through wobble pointing, using two different offset : 0.7 degree and 1.4 degree.

The assessment of run data quality is based on the Cosmic Ray rate. After the run quality selection, the total dataset amounts to 22.03 hours of stereo observation. The major factor limiting the data taking time during the observation period was poor weather conditions. After background subtraction estimated on regions symmetrically distributed around the pointing positions, we observed a total of 176.1 gamma-ray excess leading to a Li\&Ma significance of 21.3 $\sigma$.

\begin{figure}
\center
\includegraphics[width=.9\linewidth]{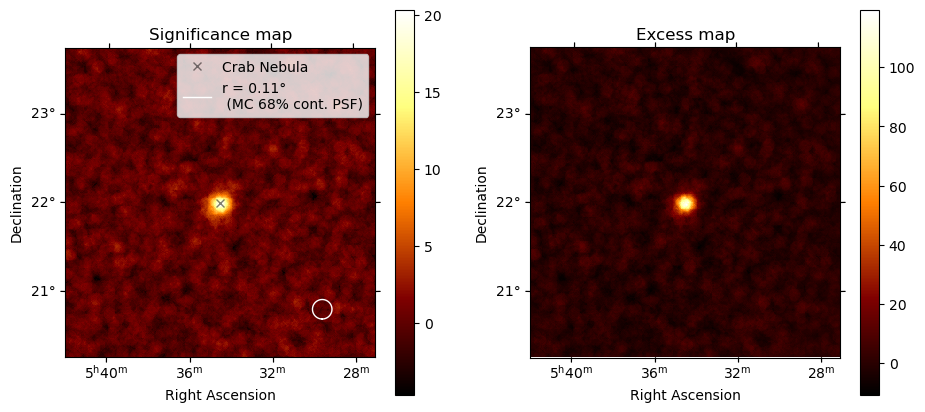}
\caption[]{Li\&Ma s´ignificance (\textit{left}) and excess (\textit{right}) sky maps derived from the Crab Nebula observation during the winter 2023/2024 campaign. The distributions are convoluted with a disk kernel of 0.07° radius.}
\label{fig:Crab}
\end{figure}

Figure \ref{fig:Crab} shows excess and significance distributions obtained with the final data set. These distributions are obtained using \texttt{gammapy} ring background method were for each position the cosmic ray background is estimated using ring background method where for each position, the cosmic ray background is estimated within a ring with inner and outer radii of 1.0 and 1.3 degrees, respectively.

\begin{figure}
\center
\includegraphics[width=.7\linewidth]{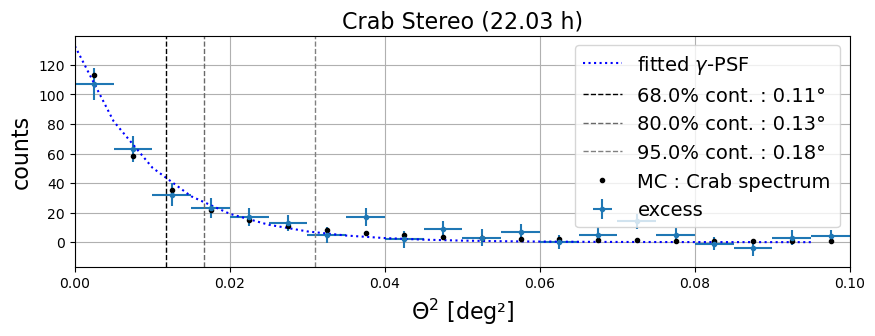}
\caption[]{$\theta^2$ distribution of the gamma-ray excess observed from the Crab nebula (\textit{blue points}) compared to the expected distribution using MC simulations weighted on the Crab spectrum for the same livetime (\textit{black points})}
\label{fig:psf}
\end{figure}

%This Crab dataset will be used to evaluate the performance of the telescopes and the analysis pipeline and compare this to the performances obtained with the MC.
This Crab dataset is crucial to evaluate the performance of the telescopes and the analysis pipeline and compare this to the performances obtained with the MC. 
Figure \ref{fig:psf} shows the $\theta^2$ distribution of the gamma-ray excess fitted with a Gaussian point spread function.

\subsection{Markarian 421}
\begin{figure}
\center{
\includegraphics[width=.35\linewidth]{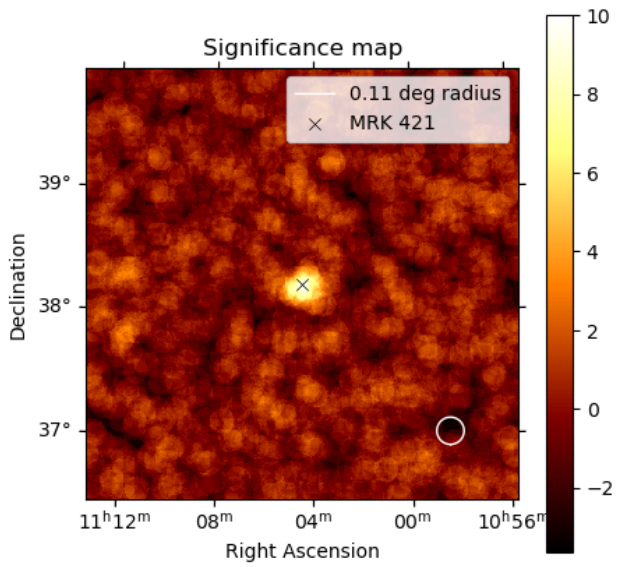}
\includegraphics[width=.35\linewidth]{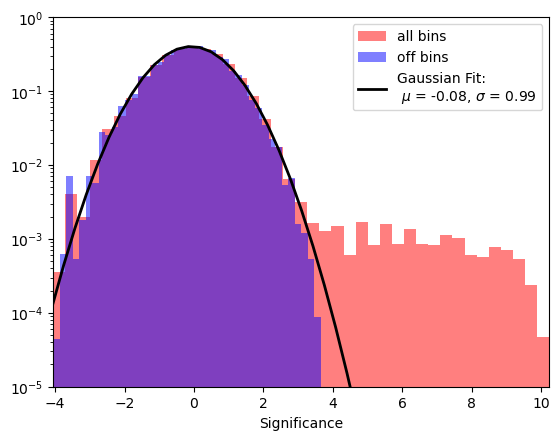}
}
\caption[]{Left panel: Li\&Ma significance map of 5.0 hours exposure on Markarian 421 during enhanced activity the 17th March 2024. Right panel : distribution of the pixel significance for all pixels (\textit{pink}) and after exclusion of the source (\textit{blue})}
\label{fig:mkn}
\end{figure}

One of the scientific objectives of the SST-1M telescope is the monitoring of blazars at TeV energies. Notably, an observation campaign was conducted on the blazar Markarian 421 during a low state of this source. Nevertheless, on 13th March 2024 the SST-1M telescopes detected a significant increase in gamma-ray flux from Markarian 421. This observation, with a livetime of approximately 3.3 hours, yielded a detection significance of 6.8 sigma, with notable excess observed up to 7 TeV and a gamma-ray flux close to the Crab.
The findings were promptly reported in Astronomer Telegram \cite{atel}.
On 17th March 2024, this increased emission was confirmed with an additional 5.0 hours of observation. Figure \ref{fig:mkn} shows significance  map 
 of the latter observation.
These observations shows the capability of the SST-1M telescope in capturing and report transient phenomena and contribute to our understanding of the high-energy astrophysical processes occurring within blazars and realated phenomena.

%\begin{figure}
%\begin{minipage}{0.5\linewidth}
%\centerline{\includegraphics[width=\linewidth]{fig/mkn_sig.png}}
%\end{minipage}
%\hfill
%\begin{minipage}{0.5\linewidth}
%\centerline{\includegraphics[width=\linewidth]{fig/mkn1d2.png}}
%\end{minipage}
%\hfill
%\caption[]{Left panel: significance map of 5.0 hours exposure on Markarian 421 during enhanced activity the 17th March 2024}
%\label{fig:mkn}
%\end{figure}

\section{Conclusion}

The winter 2023/2024 observation campaign of the Crab Nebula using the SST-1M telescopes has successfully demonstrated the expected performance of the instruments and analysis pipeline. With a total of 22.5 hours of stereo observation we achieved significant detections consistent with theoretical expectation. This validate the calibration and reconstruction processes implemented in the \texttt{sst1mpipe} pipeline.

During this first part of the commissioning phase, the SST-1M telescopes have primarily focused on the accumulation of observation on the Crab nebula and the monitoring of blazars. The next objective is to take advantage of the wide field of view of the SST-1M telescopes for the observation of extended sources. Among the other scientific objectives, the SST-1M collaboration also aims to follow alerts and targets of opportunity to contribute to the study of transient phenomena.

%
%\subsection{plop}
%
%
%Bibliography can be generated either manually or through the BibTeX
%package (which is recommanded). In this sample we
%have used \verb^\bibitem^ to produce the bibliography.
%Citations in the text use the labels defined in the bibitem declaration,
%for example, the first paper by Jarlskog \cite{ja} is 

\section*{Acknowledgments}

%The contribution of the Czech authors is supported by research infrastructure MCTA-CZ, LM2023047 MEYS, and
%Czech Science Foundation, GACR 23-05827S.
\begin{footnotesize}

The work is financed by the  Départment de Physique Nucléaire et Corpusculare, Faculty de Sciences of the University of Geneva, 1205 Genève, and the construction of the SST-1M cameras was also supported by   The Foundation Ernest Boninchi,1246 Corsier-CH, and the Swiss National Foundation (grants 166913, 154221, 150779, 143830). Funding by the Polish Ministry of Education and Science under project DIR/WK/2017/2022/12-3 is gratefully acknowledged. The Czech partner institutions acknowledge support of the infrastructure and research projects by Ministry of Education, Youth and Sports of the Czech Republic and regional funds of the European Union, MEYS LM2023047 and EU/MEYS CZ.02.01.01/00/22 008/0004632., and Czech Science Foundation, GACR 23-05827S.

\end{footnotesize}
%The Czech partner institutions acknowledge support of the infrastructure and research projects by Ministry of Education, Youth and Sports of the Czech Republic and regional funds of the European Union, MEYS LM2023047 and EU/MEYS  CZ.02.01.01/00/22\_008/0004632.
%The contribution of the  Départment de Physique Nucléaire et Corpusculare, Faculty de Sciences of the University of Geneva, 1205 Genève, and the construction of the SST-1M cameras was also supported by   The Foundation Ernest Boninchi,1246 Corsier-CH, and the Swiss National Foundation (grants 166913, 154221, 150779, 143830)
%Funding by the Polish Ministry of Education and Science under project
%DIR/WK/2017/2022/12-3 is gratefully acknowledged. The contribution of the Départment de Physique Nucléaire et
%Corpusculare, Faculty de Sciences of the University of Geneva, 1205 Genève is supported by The Foundation Ernest
%Boninchi,1246 Corsier-CH, and the Swiss National Foundation (166913, 154221, 150779, 143830).

\end{document}